\documentclass[12pt]{iopart}

\expandafter\let\csname equation*\endcsname\relax
\expandafter\let\csname endequation*\endcsname\relax

\usepackage{xcolor}
\usepackage{graphicx}
\usepackage{subfigure}
\usepackage{amsmath}
\usepackage{amssymb}
\usepackage{gensymb}
\usepackage{caption} 
\usepackage{cite}

\begin{document}

\title[Short term minutes scale statistics of sodium layer dynamics]{Short term minutes-time scale temporal variation statistics of sodium layer dynamics}

\author{Lu Feng $^1$, Kai Jin $^3$, Hong-Yang Li $^1$$^,$$^2$,  Bo-Tian Sun $^5$, Min Li $^3$, Rui-Tao Wang $^2$$^,$$^3$, Qi Bian $^4$, Chen Wang $4$, Ming Wang $^4$, Yue Liang $^4$, Zhi-Xia Shen $^1$, Yang-Peng Li $^1$, Sui-Jian Xue $^1$}

\address{$^1$ CAS Key Laboratory of Optical Astronomy, National Astronomical Observatories, Chinese Academy of Sciences, Beijing 100101, China} 
\address{$^2$ University of Chinese Academy of Sciences, Beijing 100049, China}
\address{$^3$ Institute of Optics and Electronics, Chinese Academy of Sciences, Chengdu Shuangliu 350, Sichuan, China} 
\address{$^4$ Technical Institute of Physics and Chemistry, Chinese Academy of Sciences, Beijing 100190, China} 
\address{$^5$ Department of Physics, University of California, Santa Barbara, CA 93106, USA } 
      
\eads{\mailto{jacobfeng@bao.ac.cn}}
\vspace{10pt}
\begin{indented}
\item[]June 2020
\end{indented}

\begin{abstract}
The brightness and height of the sodium laser guide star of adaptive optics could vary significantly due to the temporal dynamics of sodium column density and the mean height of sodium layer. To measure these dynamics, an independent sodium Lidar is a necessity. Without such an instrument, it is almost impossible to discern the cause of the brightness variation of laser guide star from the sodium layer's dynamics or other factors from the laser itself. For applications such as characterizing the performance of sodium laser for sodium laser guide star generation, minutes scale short term statistics of the sodium layers' abundance and height is extremely helpful for estimating the contribution of sodium layer's variation to the variation of laser guide star's brightness. In this paper, we analyzed our previous measurement of sodium layer dynamics that has been gathered in two winters, and presented the temporal variation statistics of sodium column density and mean height within minute time scale based on our measurements.
\end{abstract}

%
%
%
%
%

\section{Introduction}
\label{section: introduction}
Sodium Laser Guide Star (LGS) is becoming an indispensable component for Adaptive Optics (AO) system in recent years. In comparison with natural guide star, the artificially made sodium laser guide star could be projected to anywhere in the sky, and significantly improves the sky coverage of traditional AO system that solely depends on natural guide stars.

However, because the sodium laser guide star is generated by the spontaneous emission of sodium atoms in the mesosphere. The performances of the sodium laser guide star are bond with the variation of sodium layer which is normally located 90$\sim$110 km above sea level in the atmosphere. The abundance, or column density of sodium atoms in this layer will influence the brightness of sodium LGS, while the variation of vertical distribution of sodium layer will introduce focal anisoplanatism and change the shape of LGS spot in AO wavefront sensor and thus degrade AO system's wavefront sensing performance. It is, therefore, important to measure and study the dynamics of sodium layers on site. For the evaluation of sodium LGS performance in long-term and large geological scale, data from remote sensing satellites have been published by several sources such as \cite{Fan2007, Fussen2004, Fussen2010, Langowski2017}. High cadence sodium dynamics measurements on-site have also been conducted and reported in a number of articles \cite{Kwon1988, Gong2002, Hickson2010, Jiao2014}. However, the focus of these articles are mainly either on the overall trends of the absolute value of sodium abundance on a long time scales or on the sporadic behaviors of the sodium layer. For relatively short term application such as characterizing the performance of sodium laser for generating sodium laser guide star, which normally takes a few minutes for each test, one needs to answer the question of how much percentage of sodium column density would change within minutes time scale because the variation in these photometric results includes not only the results of the laser but also from the short dynamic change of the sodium layer.

In this paper, we attempt to use previous measurements gathered at Gao-Mei-Gu Observatory in two winters to find this answer, or at least provide a possible hint of the scale of the variation of sodium column density and sodium layer's mean height that could help similar LGS experiments evaluating laser's performances. 

In section \ref{section: data}, we will present data that we have gathered on site, and the criteria by which certain data are chosen to do statistics. In section \ref{section: method and results}, we will describe the method we used for statistics and show the results. In section \ref{section: conclusion}, conclusion is given based on our current results.  

\section{Data}
\label{section: data}
Gao-Mei-Gu observatory is located in Yunnan, China. Its latitude and longitude are $26^o42'32"N$,$100^o01'51"E$ respectively. A sodium fluorescence lidar developed by the University of Science and Technology of China was deployed at the site for monitoring the variations of column density and mean height of the sodium layer \cite{Xue2013}. The transmitter of the lidar was a tunable pulsed dye laser. The pulse energy of the laser is 60mJ. The laser's repetition rate and wavelength are 20Hz and 532nm. The central wavelength of the dye laser was firstly tuned to 589nm sodium D2 line with a sodium vapor cell, and further fine tuning was achieved by maximizing the backscattered photon return from atmospheric sodium layer. The projection direction of the laser is in the zenith. An 1.8 meter diameter telescope \cite{Rao2010} was used for collecting photons returned from the sodium layer. The collected photons were passed through a narrow band interference filter before they were finally detected by a photo-multipiler tube. The time bin length of the lidar is 100$\mu$s which corresponds to a range bin length of 150m. The number of accumulated laser shots for obtaining each lidar photo-count profile was adjusted so that for every minute we had approximately 30 measurements for column density and mean height of the sodium layer unless the returned signal was very strong so that we could shorten the accumulation time and having more measurements per minute. Main specifications of the USTC lidar system are listed in table \ref{table: lidar specification}. 

\begin{table}[]
\centering
\caption{\centering Main specifications of the USTC lidar system at Gao-Mei-Gu observatory}
\begin{tabular}{|l|l|}
\hline
\multicolumn{2}{|l|}{\textbf{Transmitter}}                        \\ \hline
wavelength (nm)                         & 589                     \\ \hline
pulse energy (mJ)                       & 60                      \\ \hline
line width (cm$^{-1}$) & 0.05                    \\ \hline
pulse width (ns)                        & 6                       \\ \hline
repetition rate (Hz)                    & 20                      \\ \hline
beam divergence (mrad)                  & 0.2                     \\ \hline
\multicolumn{2}{|l|}{\textbf{Receiver}}                           \\ \hline
diameter (m)                            & 1.8                     \\ \hline
field of view (mrad)                    & 1.0                     \\ \hline
\multicolumn{2}{|l|}{\textbf{Receiver Filter}}                    \\ \hline
center wavelength (nm)                  & 589                     \\ \hline
bandwidth (nm)                          & 1.0                     \\ \hline
peak transmission                       & $\sim$50\% \\ \hline
\end{tabular}
\label{table: lidar specification}
\end{table}

\begin{figure}
    \centering
    \includegraphics[width=0.8\linewidth]{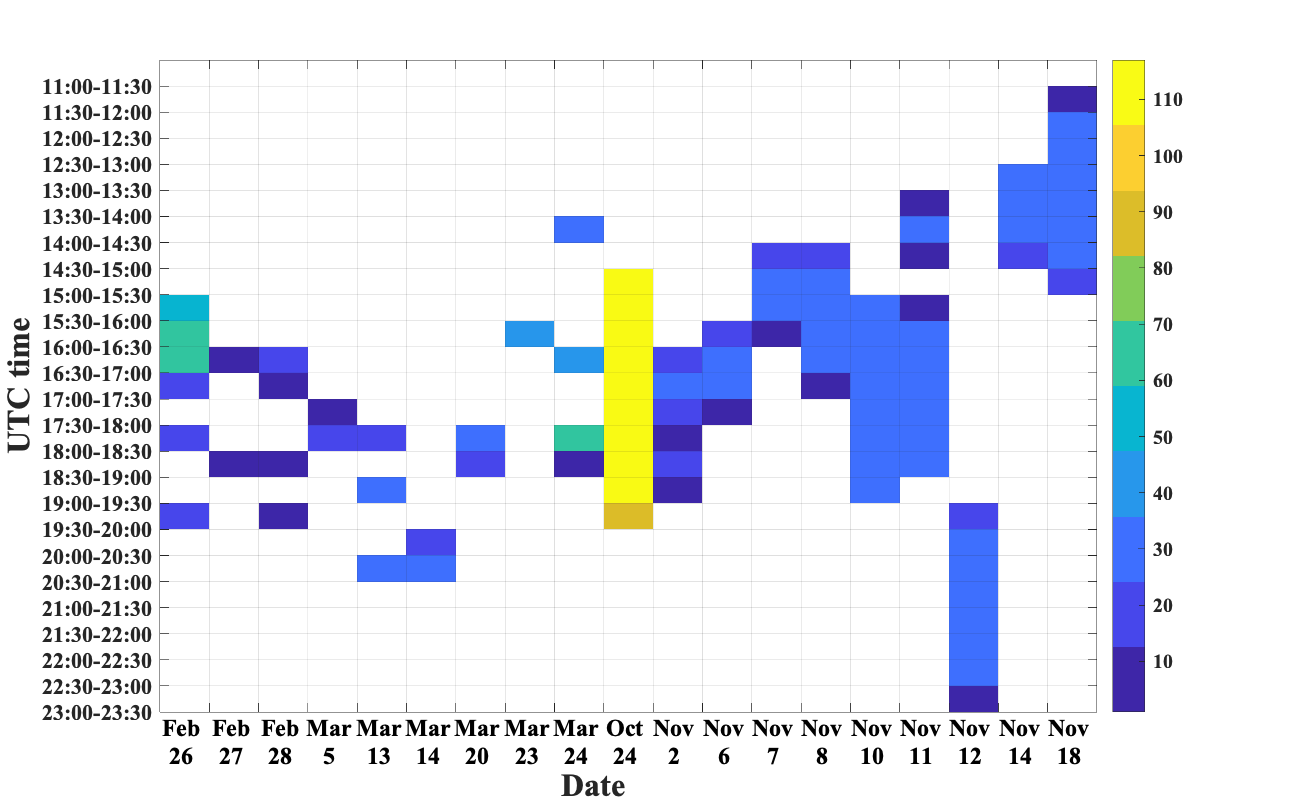}
    \caption{The sodium lidar's night time data availability. X axis is the date of the measurement. Y axis is the UTC time for every half hour. Different colors in the grids represent the amount of data gathered during half-hour period.}
    \label{fig: data availability}
\end{figure}

\begin{figure}[htbp]
    \centering
    \subfigure[sodium column density]{
        \begin{minipage}[t]{0.45\linewidth}
        \centering
        \includegraphics[width=\linewidth]{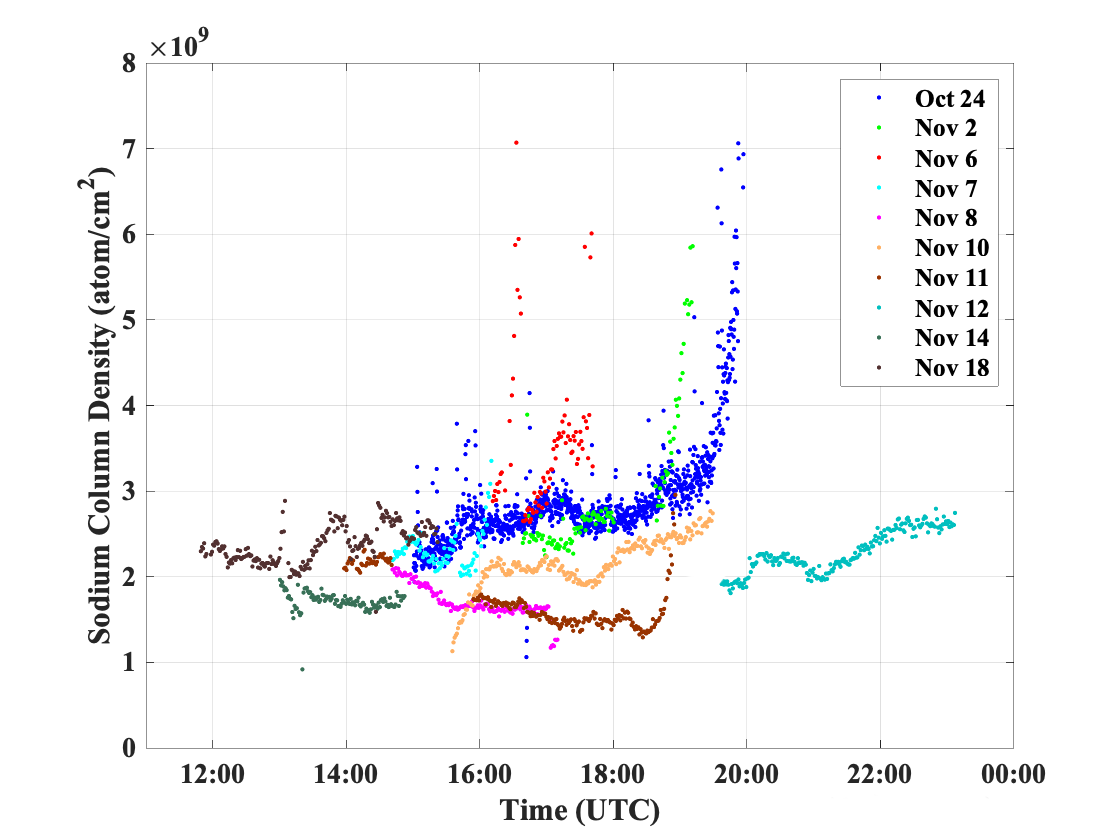}
        \end{minipage}
    }
    \subfigure[sodium layer's mean height]{
        \begin{minipage}[t]{0.45\linewidth}
        \centering
        \includegraphics[width=\linewidth]{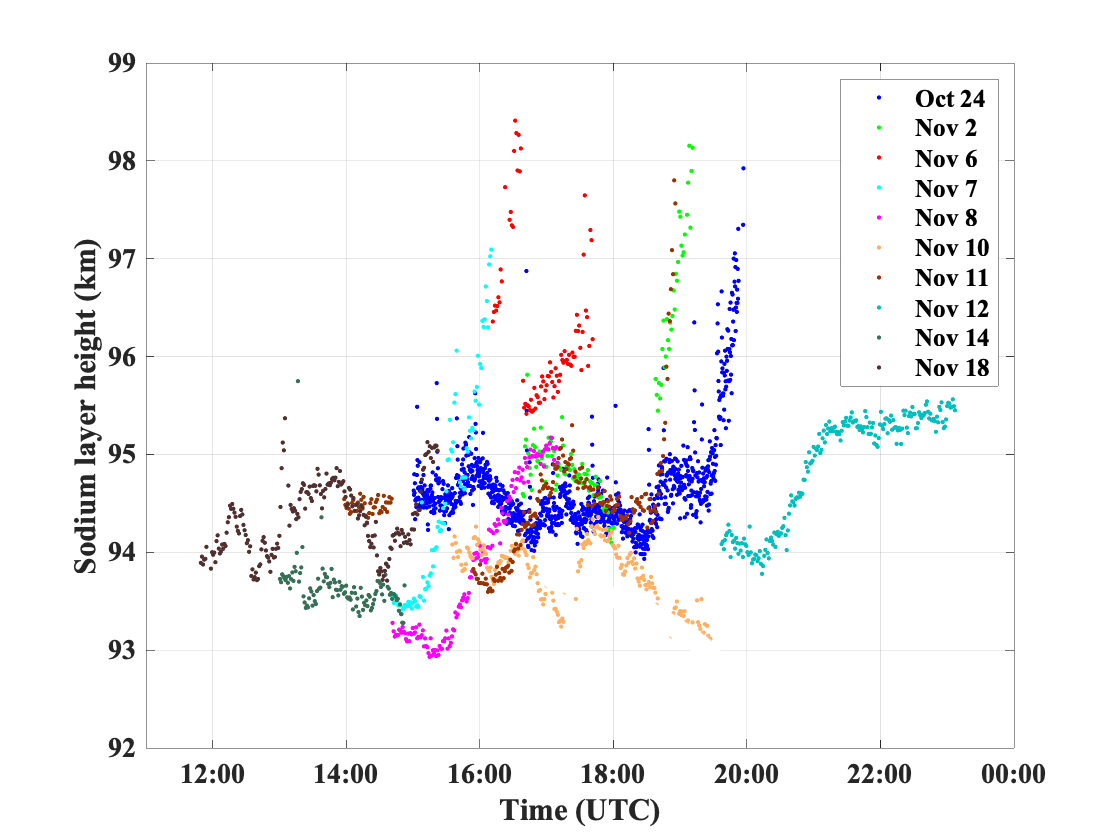}
        \end{minipage}
    }
    \caption{Data for sodium column density (a) and sodium layer's mean height (b) gathered during the second phase. Coloring for different dates are the same in both figures.}
    \label{fig: raw data}
\end{figure}

Observation of the sodium layer were conducted separately in two phases in 2013. The first phase was from late February to the end of March. The second phase started from late October and ended by late November. A total of 19 nights' data were collected. Figure \ref{fig: data availability} presented the data availability for all measurements. During the first phase, the weather condition is not favorable. The continuity and density of data for the first phase was therefore not on par with the second phase as shown in figure \ref{fig: data availability}. Due to this reason, further statistical analysis for sodium layer's dynamics were based only on data obtained in the second phase, which included 10 nights of data. Figure \ref{fig: raw data} showed these measurements for both the sodium column density and sodium layer's mean height. 

Because that the sodium Lidar would drop bad values if signal to noise ratio was low and also that some of the occasional spikes could be contributed by the sporadic behaviors of sodium layer that could actually happen during a laser guide star, all data produced by the Lidar were included in our statistics.

\section{Method and results}
\label{section: method and results}

In this analysis, we only focused on the variation of sodium dynamics in short term minute time scale. For typical sodium laser guide star tests such as measuring the brightness variation against different changing laser parameters like output power level, polarization or modulation depth of $D_{2b}$ re-pumping, etc., the duration of such single test is normally in the range of 5 to 20 minutes. The relationship between the brightness or the photon flux $\Phi$ of sodium laser guide star and the sodium column density $C_{Na}$ can be described with equation \ref{equation: lidar equation}, where $L$ is the mean height of the atmospheric sodium layer, $P$ is the laser power, $T$ is the atmospheric transparency, $\theta$ is the laser launch zenith angle, and $X=1/(sec(\theta))$.  

\begin{equation}
  \Phi = \frac{S_{ce}\cdot P \cdot (T)^{2X} \cdot C_{Na} \cdot X}{L^{2}}  
\label{equation: lidar equation}
\end{equation}

The variation of photon flux due to the change of height $\Delta L$  and sodium column density $\Delta C$ can then be described as,

\begin{equation}
    \begin{split}
        \frac{\partial \Phi}{\partial C}{/\Phi} = \frac{\Delta C}{C} \approx \frac{\Delta C}{\bar C}\\
        \frac{\partial \Phi}{\partial L}{/\Phi} = -\frac{2\Delta L}{L} \approx -\frac{2\Delta L}{\bar L}\\
    \end{split}
    \label{equation: variations}
\end{equation}

Therefore, in order to isolate the variation of brightness of the sodium laser guide star from the variation of sodium layers dynamics for such tests, we need to analyze the statistics of sodium column density and the mean height of sodium layer within a corresponding time duration. 

A moving time window with a  window size of fixed duration of $\Delta T$ were used for picking out data set for statistics calculation. This time windows was then applied from the beginning of each night's data, and was shifted for every measurements after mean and standard deviation (STD) for previous set were calculated until the right side of the window reached the last measurement of the day. 

\begin{figure}
    \centering
    \includegraphics[width=\linewidth]{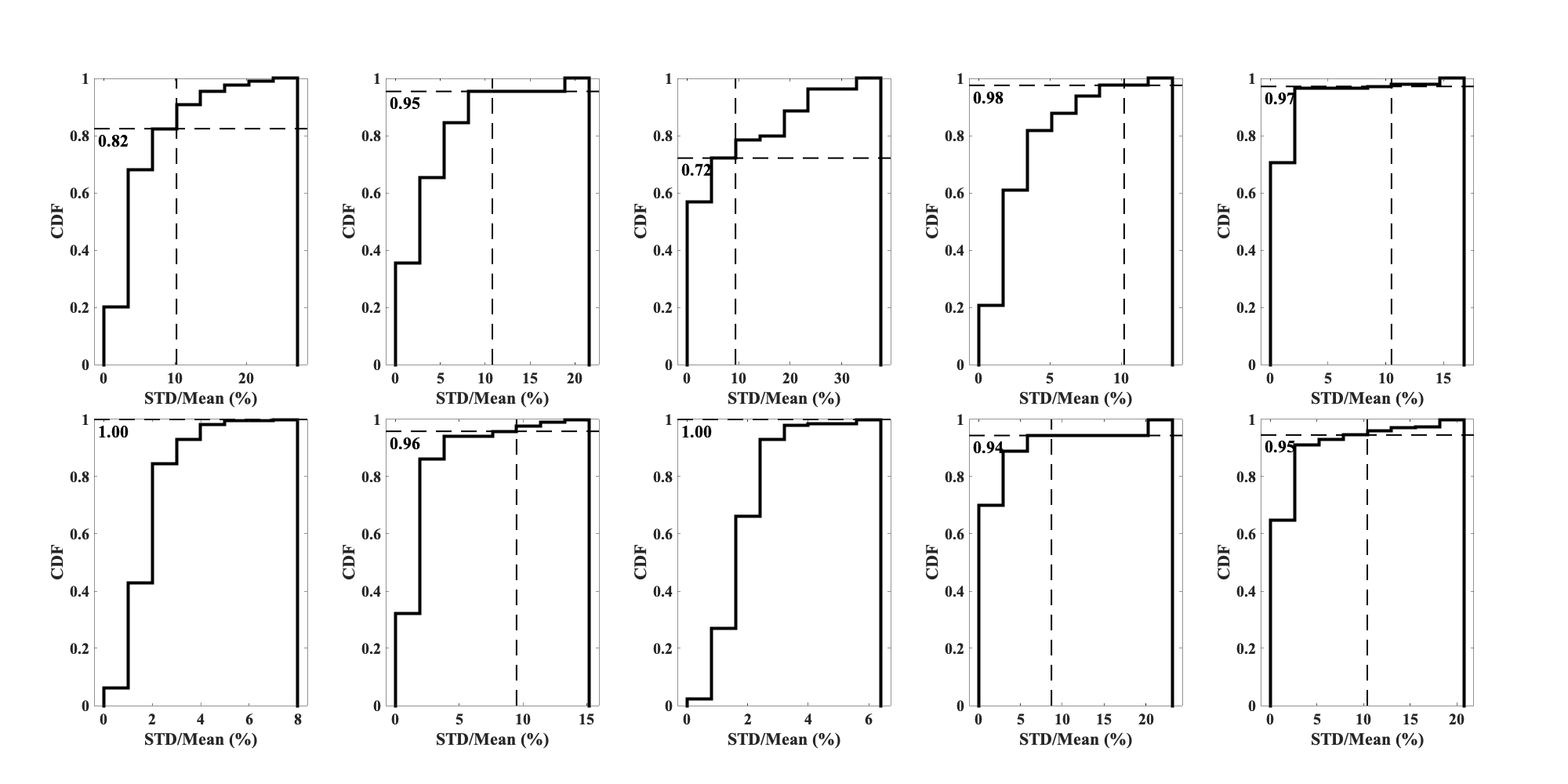}
    \caption{CDF of Sodium column density statistics with a 10 minutes time window. The upper panels represent data for October 24th, November 2nd, 6th, 7th and 8th from left to right. The lower panels represent data for November 10th, 11th, 12th, 14th and 18th. The x axis for each panel is the ratio of standard deviation to mean value. The y axis is the cumulative distribution function in fraction. The vertical dash line in each panel shows where the ratio is $10\%$ and the corresponding CDF value is shown by the horizontal line.}
    \label{fig: 10 minutes sodium column density cdf}
\end{figure}

\begin{figure}
    \centering
    \includegraphics[width=\linewidth]{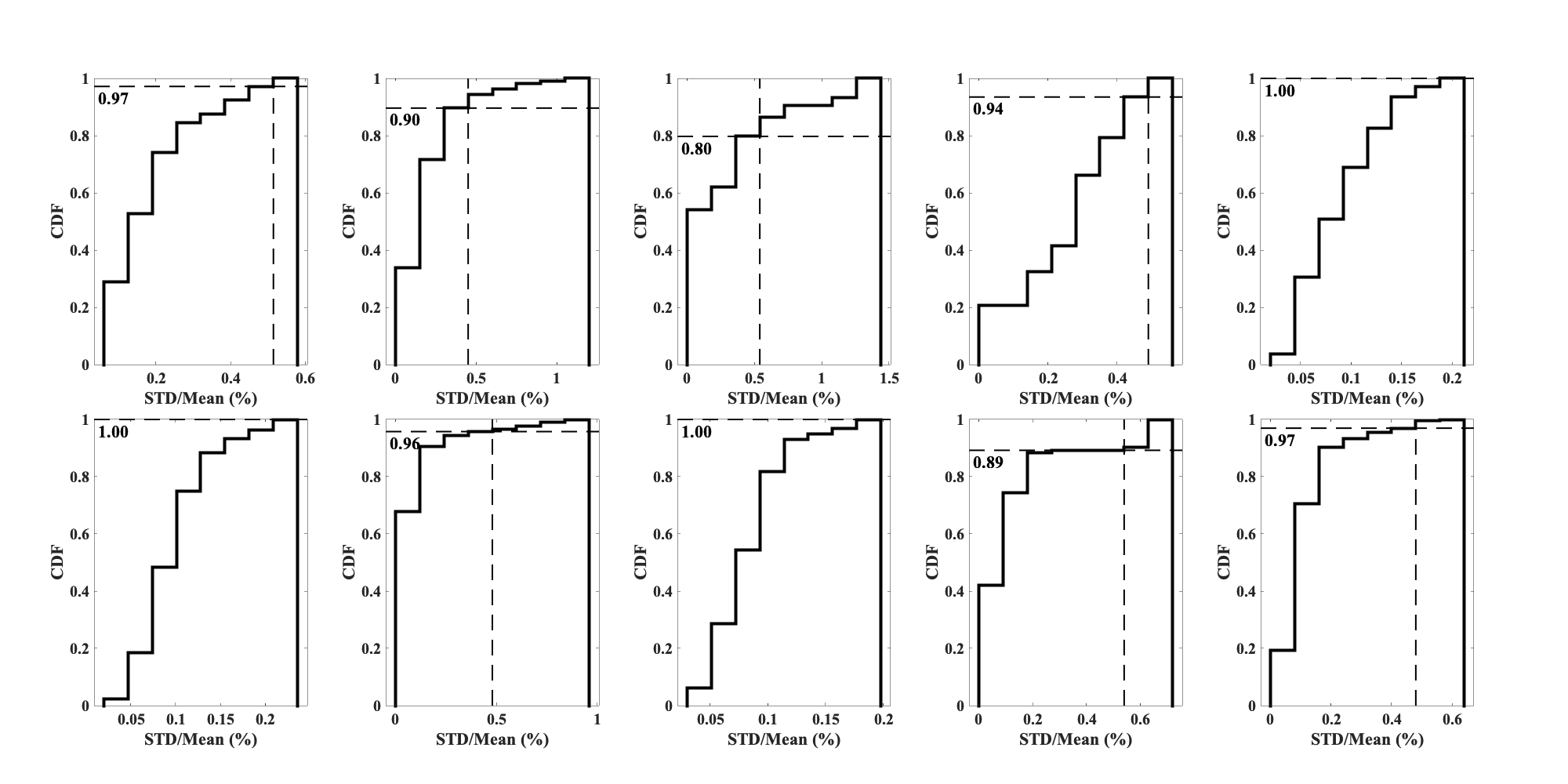}
    \caption{CDF of Sodium layer's mean height statistics with a 10 minutes time window. The upper panels represent data for October 24th, November 2nd, 6th, 7th and 8th from left to right. The lower panels represent data for November 10th, 11th, 12th, 14th and 18th. The x axis for each panel is the ratio of standard deviation to mean value. The y axis is the cumulative distribution function in fraction. The vertical dash line in each panel shows where the ratio is $1\%$ and the corresponding CDF value is shown by the horizontal line.}
    \label{fig: 10 minutes sodium layer height cdf}
\end{figure}

The ratio of standard deviation $\sigma_i$ to mean value $\mu_i$ for each data set $i$ would represent the relative variation of sodium layer within $\Delta T$ at the time of the data set $i$. The Cumulative Distribution Function (CDF) of this ratio for each night would indicate the possibility that the column density or the mean height changed by certain percentage within $\Delta T$ during that night. Figure \ref{fig: 10 minutes sodium column density cdf} shows the 10 minutes statistics of sodium column density for each night, and figure \ref{fig: 10 minutes sodium layer height cdf} shows the 10 minutes statistics for sodium layer height. 

With this method, we could analyze for each of these 10 nights, the possibility that the percentage of variation is lower than a certain level within different duration of time. For our laser testing scenario, we arbitrarily set this level to $10\%$ for column density and $0.5\%$ for sodium layer height according to our laser characterization test requirements. Statistics results with different time windows as shown in figure \ref{fig: 10 minutes sodium column density cdf} and \ref{fig: 10 minutes sodium layer height cdf} were also provided on our website which could be used to derive probability with other levels. 

\begin{figure}[htbp]
    \centering
    \subfigure[sodium column density]{
        \begin{minipage}[t]{0.45\linewidth}
        \centering
        \includegraphics[width=\linewidth]{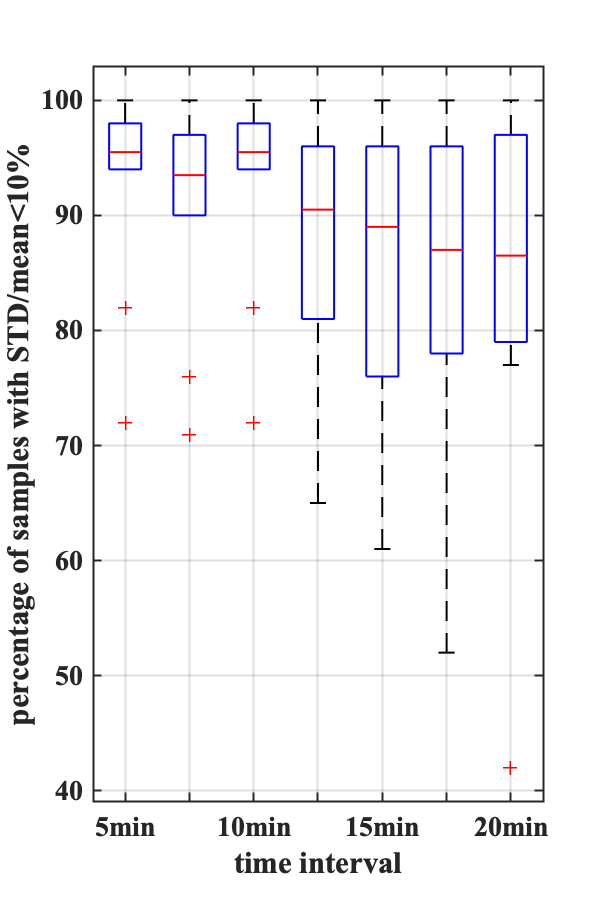}
        \end{minipage}
    }
    \subfigure[sodium layer's mean height]{
        \begin{minipage}[t]{0.45\linewidth}
        \centering
        \includegraphics[width=\linewidth]{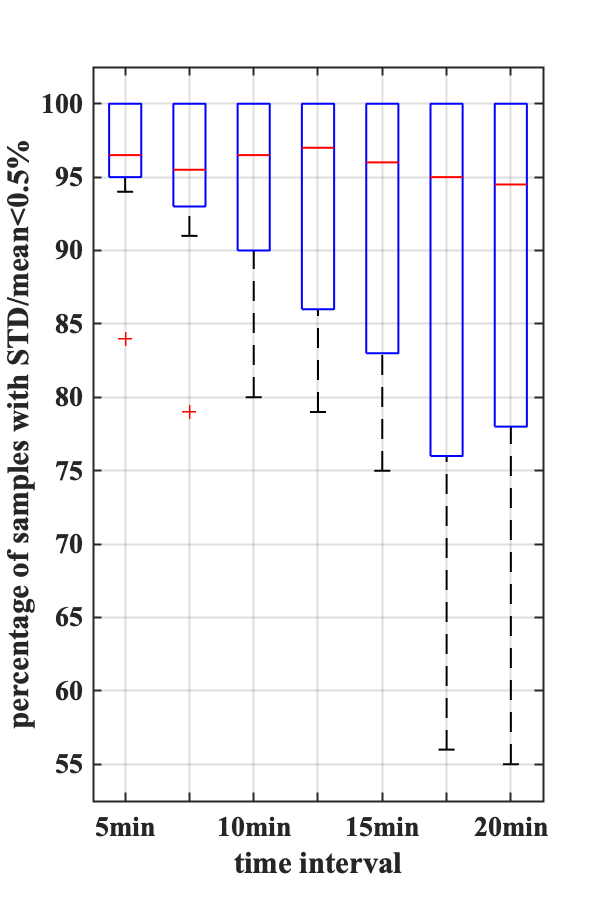}
        \end{minipage}
    }
    \caption{Percentages of samples or approximate possibilities with ratio of STD to mean less than $10\%$ and $0.5\%$ for sodium column density and mean height with different time window size. On each box, the central mark indicates the median value, and the bottom and top edges of the box indicate 25th and 75th percentiles. Wiskers represents extreme values, and red crosses are outliers.}
    \label{fig: final statistics}
\end{figure}

Figure \ref{fig: final statistics} shows the results of these statistics for both sodium column density (left) and sodium layer's mean height (right). 
 
\section{Conclusion}
\label{section: conclusion}
The motivation of this paper is to help sodium laser guide star performance test to determine a suitable time duration for single test set so that even without a LIDAR one could claim if the brightness variation of sodium laser guide star is due to sodium layer or laser with certain confidence. Therefore, based on our previous measurements, we analyzed the minute time scale statistics of sodium column density and sodium layer's mean height considering such test sets are normally in the range of 5 to 20 minutes. 

Figure \ref{fig: final statistics} shows the approximate possibilities of sodium column density and sodium layer's mean height to be within given range aggregating all 10 nights' statistic results. As shown in the figure, with an larger time window, the confidence for claiming the variation of column density or mean height to be within $10\%$ or $0.5\%$ is lower, and the scattering is also worsening which represent differences of sodium dynamics between these 10 nights. If the duration is limited within 10 minutes, there is high probability ($>90\%$) that the ratio of sodium column density's STD to mean value is less than $10\%$. For the mean height of the layer, not only the variation is small with a ratio of STD/mean less than $0.5\%$, but also that for a duration less than 10 minutes, the probability to have a small variation is also high. If the time window extends to a longer time, the confidence to have such claims, especially for column density drops obviously.

Equation \ref{equation: variations} shows that the variation of brightness of the sodium laser guide star are in linear relationship with variations of column density and mean height. However, comparing between these two factors of sodium layer with the aforementioned results, the influence of column density on the brightness of sodium laser guide star is much more significant than mean height. 

In light of these, for sodium laser guide star performance test without the aid of LIDAR, if the laser guide star's brightness is expected to be changed higher than $10\%$ with a changing parameter of the laser, a test set with duration shorter than 10 minutes would be sufficient. The test duration needs to be shorter if the precision requirement is more stringent.

\section{Acknowledgement}
\label{section: ackowledgement}
The research is partly supported by the Operation, Maintenance and Upgrading Fund for Astronomical Telescopes  and  Facility  Instruments,  budgeted  from  the  Ministry  of Finance of China (MOF) and administrated by the Chinese Academy of Sciences (CAS).

\end{document}